\documentclass[letter]{aa}
\usepackage{txfonts}
\usepackage{natbib}
\usepackage{graphicx}
\bibpunct{(}{)}{;}{a}{}{,}
\begin{document}

\title{The unusual Nova Cygni 2006 (V2362 Cyg)}
\author{S. Kimeswenger\inst{1}
\and S. Dalnodar\inst{1} \and A. Knapp\inst{1} \and J.
Schafer\inst{1} \and S. Unterguggenberger\inst{1} \and S.
Weiss\inst{1}} \institute{$^{\rm 1}$Institut f\"ur Astro- und
Teilchenphysik, Universit\"at Innsbruck, Technikerstra{\ss}e 25,
6020 Innsbruck, Austria}

\date{Received 10 October 2007 / Accepted 11 January 2008}

\abstract {Optical nova lightcurves often have structures, such as
rapid declines and recoveries, due to nebular or dusty phases of the
ejecta. Nova Cygni 2006 (V2362 Cyg) underwent an unusual brightening
after an early rapid decline.  The shape of the lightcurve can be
compared to that of V1493 Aql, but the whole event in that case was
not as bright and only lasted a couple of weeks.  V2362 Cyg had a
moderately fast decline of $t_2 \,=\,9.0\pm0.5$ days before
rebrightening, which lasted 250 days after maximum.} {We present an
analysis of our own spectroscopic investigations in combination with
American Association of Variable Star Observers (AAVSO) photometric
data covering the whole rebrightening phase until the return to the
final decline.} {We used the medium resolution spectroscopy obtained
in ten nights over a period of 79 nights to investigate the change
of the velocity structure of the ejecta. The publicly available
AAVSO photometry was used to analyze the overall properties and the
energy of the brightening.} {Although the behavior of the main
outburst (velocity, outburst magnitude, and decline timescales)
resembles a ``normal'' classical nova, the shell clearly underwent a
second fast mass ejecting phase, causing the unusual properties. The
integrated flux during this event contributes $\approx$40\% to the
total radiation energy of the outburst. The evolution of the
H$_\alpha$ profile during the bump event is obtained by subtracting
the emission of the detached shells of the main eruption by a simple
optically thin model. A distance of $D \approx
7.5^{+3.0}_{-2.5}$~kpc and an interstellar extinction $E(B-V) =
0\fm6\pm0\fm1$ was also derived.} {}


%\titlerunning{<short title>}
%\authorrunning{<name(s) of author(s)}

\maketitle

\section{Introduction}

NOVA Cygni 2006 (V2362 Cyg) was discovered April 2, 2006 by H.
Nishimura as a 10\fm5 object \citep{discover}. The early spectra,
obtained $\approx 0\fd6$ prior to the maximum in $V$, showed a
rather flat continuum, on which the H$_\alpha$ line shows a clear
P-Cyg profile. The FWHM of the emission was 500 km s$^{-1}$ and the
absorption minimum was 700 km s$^{-1}$ blueshifted
\citep{IAUC8698_1}. Within a day ($\approx -0\fd2$), the profiles
showed two absorption features at --880 and --1\,730~km~s$^{-1}$
\citep{early_spec}. Based on spectra obtained on April 13, 2006
($\approx 6\fd7$), \citet{siviero} classified the nova as 'Fe-II'
and found a very broad and structured profile with FWZI of
3\,750~km~s$^{-1}$ and FWHM of 1\,800~km~s$^{-1}$ from the Balmer
emission lines.

With the exact position found by \citet{position},
\citet{progenitor} identified the progenitor as an emission line
star in the IPHAS H$_\alpha$ survey. With progenitor magnitudes of
$r'=20\fm3\pm0\fm05$ and $i'=19\fm76\pm0\fm07$ the outburst
amplitude was about 12 magnitudes.

In July 2006 ($\approx 100^{\rm d}$) the first unusual behavior was
reported by \citet{no_ir_dust}. The nova, after having already
declined by 4$^{\rm m}$, still showed only low excitation lines. The
target brightened \citep{munari_c} from August 2006 ($\approx
130^{\rm d}$) until beginning of December 2006 ($\approx 240^{\rm
d}$) . \citet{goranskij_1} found an optical period of 0\fd2070 and
an increase of the emission line width during this phase. They also
pointed out similarities  to the lightcurve of V1493~Aql (Nova Aql
1999a). The second maximum in V1493~Aql had a shorter duration of 2
weeks and a smaller amplitude of only $\approx$0\fm75. The duration
in V2362 Cyg was 6 months with an amplitude of 2\fm1. At the
beginning of December 2006 the brightness dropped by 2\fm5 over a
period of two weeks to a value expected from the extrapolation of
the early decay. The spectra also changed rapidly
\citep{nebular_phase} and reached a typical nebular phase at the end
of December 2006 \citep{munari_d}. The formation of very hot dust,
$T_{\rm d} \approx 1\,410~K$, began before December 12, 2006
($\approx 250^{\rm d}$) \citep{ir_dec,ir_late}. In May 2007
($\approx 396^{\rm d}$), \citet{ir_may07} reported continuing dust
emission but at a lower temperature, $T_{\rm d} < 520\,$K.

\citet{x-ray} also detected the source in X-rays on October 14, 2006
($\approx 191^{\rm d}$), and reported that the spectrum is harder
than expected from supersoft X-ray binaries. On May 5, 2007
($\approx 394^{\rm d}$), \citet{XMM} reported that the XMM-Newton
spectra were poorly fitted with a single component absorbed model.
They fitted a two-temperature thermal plasma model with a low and a
high temperature of 0.2~keV and 2.3~keV, respectively.

Here we report the results of a spectral monitoring throughout the
main brightening phase and further on until the target reached its
final decline.

\section{Data}
We obtained the spectroscopic data in ten nights in the period from
early October 2006 until the end of December 2006 at the 60~cm
telescope of the University of Innsbruck. We used the 10C slit
spectrograph with a 50~$\mu$m slit and the 240~l/mm grating was
used. The sampling at the Kodak 0400 CCD was 0.26~nm/pixel and the
resolution, measured as FWHM of the night sky lines, was 0.6~nm. The
combined frames typically covered 400-900nm each night. The
observational log is given in Tab.~\ref{obs}. For a more detailed
description of the instrument see \citet{teleskop} and references
therein.

\begin{table}[ht]
\caption{Observing log of spectroscopy and wavelength range
covered.}\label{obs}
\begin{tabular}{l c l c}
date & wavelength   & observers & ~day$^*)$\\
 &  range [nm] & \\
\hline

10$^{\rm th}$ Oct. & 400.9-573.5  & KS & 187\\
  &  591.4-791.6 & \\
16$^{\rm th}$ Oct.  & 400.0-900.0 & SD, JS, SU, KS& 193\\
17$^{\rm th}$ Oct.  & 400.0-900.0 & KS& 194\\
15$^{\rm th}$ Nov.  & 400.0-893.5 & SD, AK, JS, SU, SW, KS&223 \\
%%16$^{\rm th}$ Nov.  & & SD, AK, JS, SU, SW, KS& 228\\
20$^{\rm th}$ Nov.  & 400.0-736.7 & KS& 228\\
$\,\,\,$7$^{\rm th}$ Dec.  & 400.0-900.0& KS& 245\\
13$^{\rm th}$ Dec.  & 400.0-900.0 & SD, AK, JS, SU, SW, KS& 251\\
14$^{\rm th}$ Dec.  & 400.0-900.0 & SD, SU& 252\\
%%20$^{\rm th}$ Dec.  & & KS& \\
23$^{\rm th}$ Dec.  & 537.5-740.0& KS& 261\\
27$^{\rm th}$ Dec.  & 400.0-740.0& KS& 265\\

\hline
\end{tabular}
\newline
\noindent{\scriptsize $^*)$ days since maximum in AAVSO data
April 06, 2006 09:33 UT (JD 2453831.8983)}\\
\noindent{\scriptsize SD = Silvia Dalnodar, AK = Andreas Knapp, JS =
Josef Schafer,
\newline SU = Stefanie Unterguggenberger, SW = Sarah Weiss, KS =
Stefan Kimeswenger}
\end{table}

We obtained the flux calibration using Vega (HR~7001), HR 8252 and
HR 8079. The calibration was verified by folding the resulting
spectra with standard $V$ and $I_{\rm C}$ filter curves. The zero
points were 24\fm23 and 23\fm12 in $V$ and $I_{\rm C}$,
respectively. The rms of the photometric nights was 0\fm056. The
nights of November 20, 2006 December 7 and 27, 2006 were not
photometric. We corrected the spectra to fit to the published
photometries for those three nonphotometric nights. The resulting
zero points are consistent with the observational average.

\begin{figure} \centering\includegraphics[width=88mm]{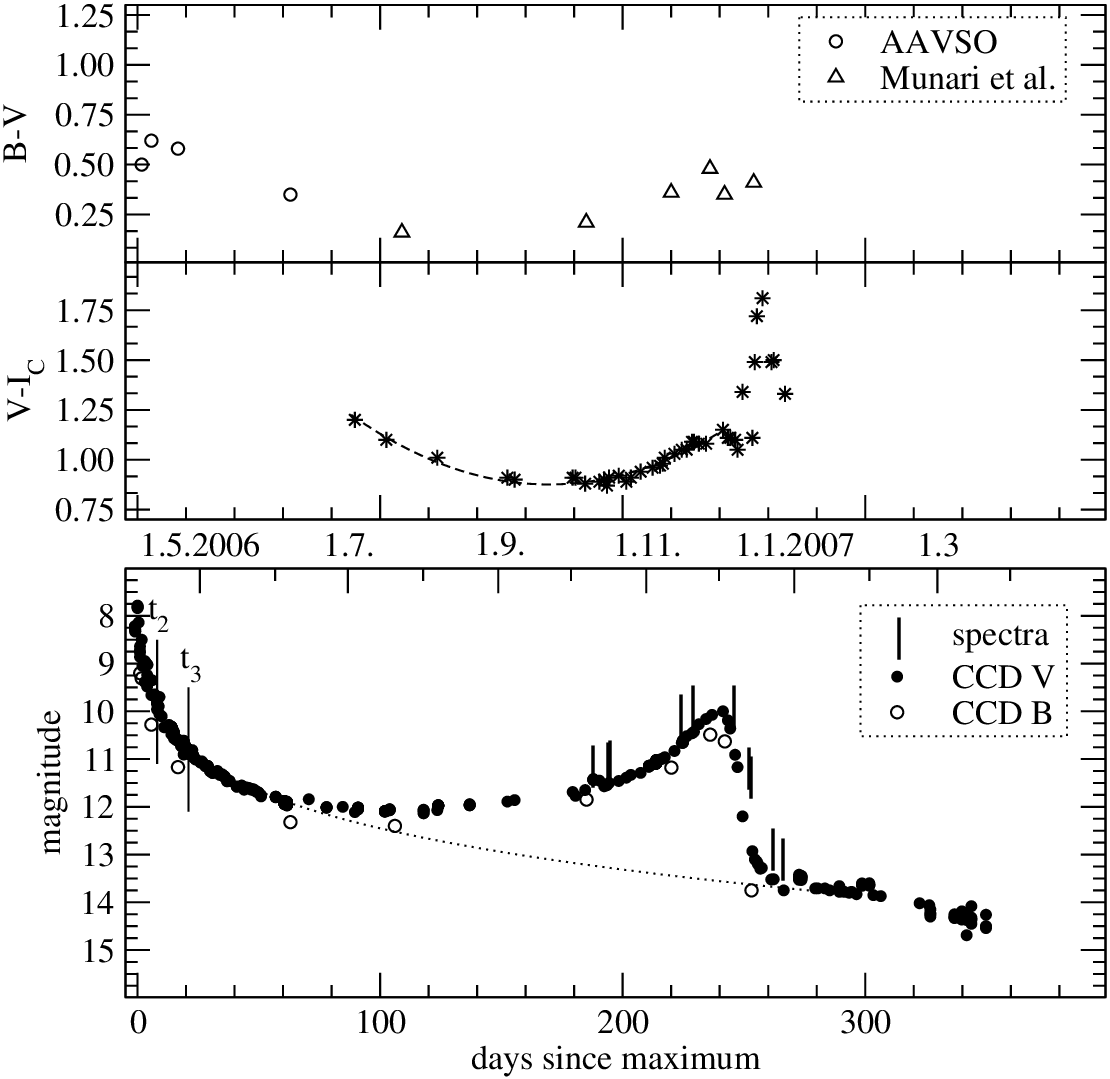}
\caption{The photometric data of V2362 Cyg from April 2006 to March
2007. The fit of a power law decline to the data obtained the first
60 days is extrapolated to show the extension for the time after day
260 (dotted line). The dates of our spectral observations and of
$t_2$ and $t_3$ used for the calculations in this paper are
indicated.} \label{lightcurve}
\end{figure}

The photometric data was obtained from the AAVSO data base
\citep{aavso} and various circulars (Munari et al 2006a,b,c,d;
Goranskij et al. 2006). We used only CCD-based photometry with
filters for the lightcurves shown in Fig.~\ref{lightcurve}. The
first values with CCDs and a $V$ band filter were reported April 5,
2006 6:56 UT (JD 2453830.7891). The maximum value of 7\fm8 was
reached April 6, 9:33 UT (JD 2453831.8983). We used this value as a
``reference zero point'' for all calculations. The decline derived
from there was $t_2$ = 9\fd0$\pm0\fd5$ and $t_3$ = 21\fd0$\pm0\fd5$.
Clear glass plate and visual observations were omitted due to the
high scatter. The decline for the first 60 days closely resembles a
power law. In addition, this fit to the early data points gives an
amazingly perfect extrapolation for the time after day 260. The
$V-I_{\rm C}$ color evolution shows that the temperature of the
quasi-photosphere varied only between 4\,250 and 4\,900 K during the
bump period, consistent with our spectra. At maximum, the
temperature was $\approx$ 8\,100\,\,K \citep{early_spec}.

\section{Absolute Magnitude, Reddening and Distance}
Nova distances can be derived from a comparison between their
intrinsic and observed luminosity and therefore require that they
are corrected for the effect of interstellar extinction.
\citet{vdBY} found that after correction for reddening, the
intrinsic color of novae two magnitudes below maximum, i.e. at time
$t_2$, is fairly constant $(B-V)_0 = -0\fm02$, with a typical
individual error (rms) of 0\fm12. Although the coverage of $B$
values is sparse in our case, assuming a smooth behavior of $B$ like
that in $V$ and $t_2$ = 9$\pm0\fd5$ we have been able to derive
$(B-V) @ t_2 = 0\fm58\pm0\fm03$ (see Fig. \ref{lightcurve}) and thus
$E(B-V) = 0\fm6\pm0\fm1$. Our value is in agreement with $E(B-V) =
0\fm56$ from the interstellar Na I absorption lines \citep{siviero},
$0\fm59$ obtained from infrared observations \citep{ir_ebv}, $0\fm6$
from fluorescently excited O I lines \citep{no_ir_dust}, and
$0\fm65\pm0.05$ obtained using the progenitors colors and assuming a
thick accretion disk spectrum \citep{progenitor}. Thus, an $E(B-V)$
of $0\fm6\pm0\fm1$ ($A_{\rm V} = 1\fm9 \pm 0\fm3$) is assumed
throughout the paper.

Several methods for distance determinations of novae are suggested
in the literature. The empirical relation between absolute magnitude
at maximum and the rate of decline \citep{MMRD} leads to $M_V
\approx -8\fm78 \pm 0\fm60$ (using the conservative 3 $\sigma$
boundary of the relation).  With a  corrected peak of $V_0({\rm
MAX}) =
5\fm93$, we derive a distance of 8.8$^{+9}_{-4}$~kpc. \\
The standard candle method by \citet{standard_1} in the
\citet{warner} calibration $M_V = -5\fm44$ fifteen days after visual
maximum yields  (corrected $V_{\rm 15^{\rm d}} = 8\fm55$) a distance
of 6.4$^{+4.5}_{-2.5}$~kpc. Finally, assuming a luminosity above
Eddington at maximum as
described in \citet{Boni00} we obtained 7.2$^{+5}_{-3}$~kpc.\\
We use a nonweighted mean of $D \approx 7.5^{+3.0}_{-2.5}$~kpc.

\section{The Spectra}
\begin{figure}[ht]
\centerline{\includegraphics[width=88mm]{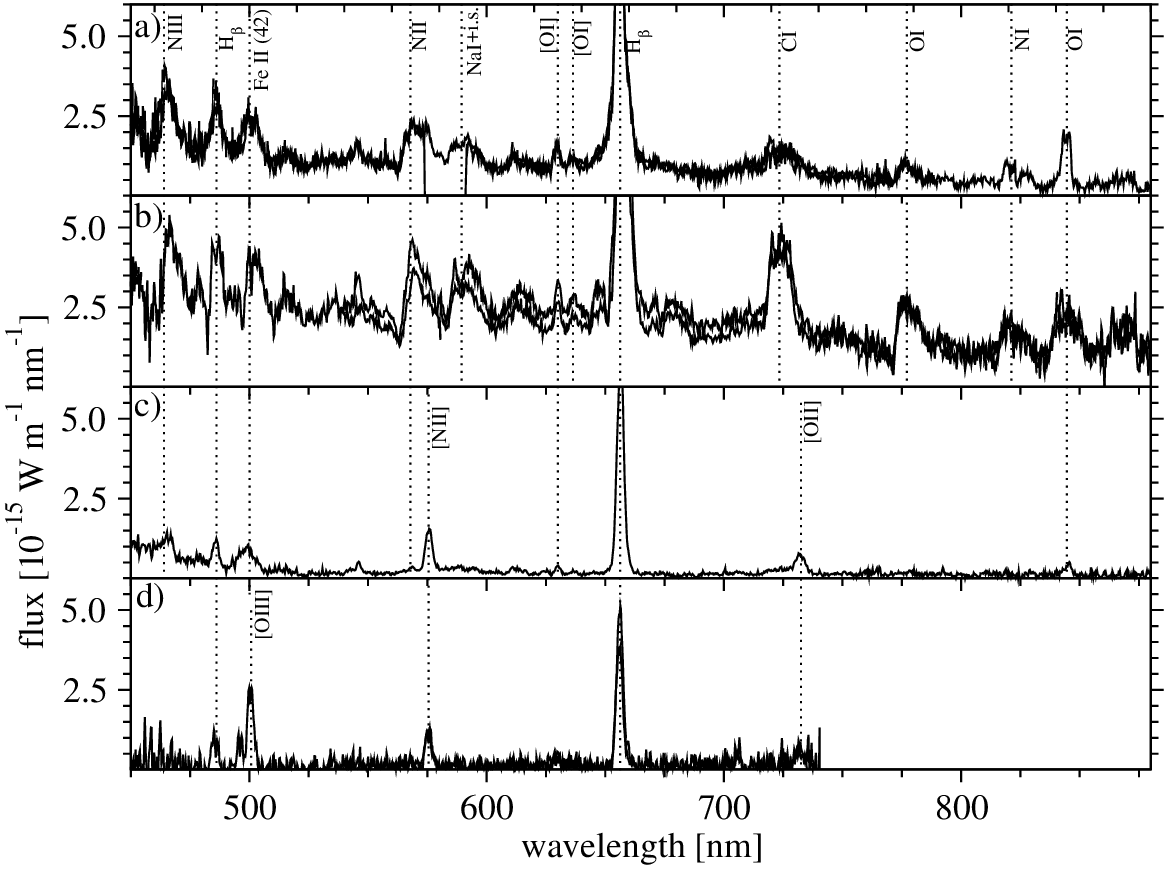}} \caption{The
flux calibrated spectra: a) October 10 and 16/17, 2006 - start of
the rebrightening; The spectra are nearly completely overlapping. b)
November 15/20 and December 7, 2006 - at 2$^{\rm nd}$ maximum; c)
December 13/14, 2006 - transition phase; d) December 23 and 27, 2006
- nebular phase; only H$_\alpha$ fades while the prominent forbidden
lines are constant. }\label{red}
\end{figure}
Four types of spectra can be seen (Fig. \ref{red}). The spectra of
October 2006 are dominated by the lines that have been visible
throughout the whole normal decline starting in April 2006. The
lines show the narrow structure, as reported by \citet{siviero}.
Only H$_\alpha$ starts to form a red shoulder of high-velocity
material. The spectra taken on October 16 and 17, 2006 showed no
differences and thus were averaged to improve S/N for the further
analysis. The spectra from November 2006 and December 7, 2006 are
dominated by massive emissions of atoms at low excitation states.
All show the wide red extension of the emission discussed in more
detail for the hydrogen lines below. Most of the prominent lines
show a fast blue P-Cygni absorption component at
$\ge$~2\,300~km~s$^{-1}$. The data taken on November 15 and 20, 2006
again have been combined into an average spectrum. By mid-December
2006, a transition spectrum emerged. Most of the nonforbidden lines
have faded and the width of the Balmer lines had decreased to their
values before October 2006. Forbidden lines started to grow and
finally dominated (except for H$_\alpha$) the nebular spectra taken
at the end of December. All spectra may be obtained electronically
from the authors.

\section{The Evolution of the H$_\alpha$ line}

\begin{figure}[ht]
\centerline{\includegraphics[width=88mm]{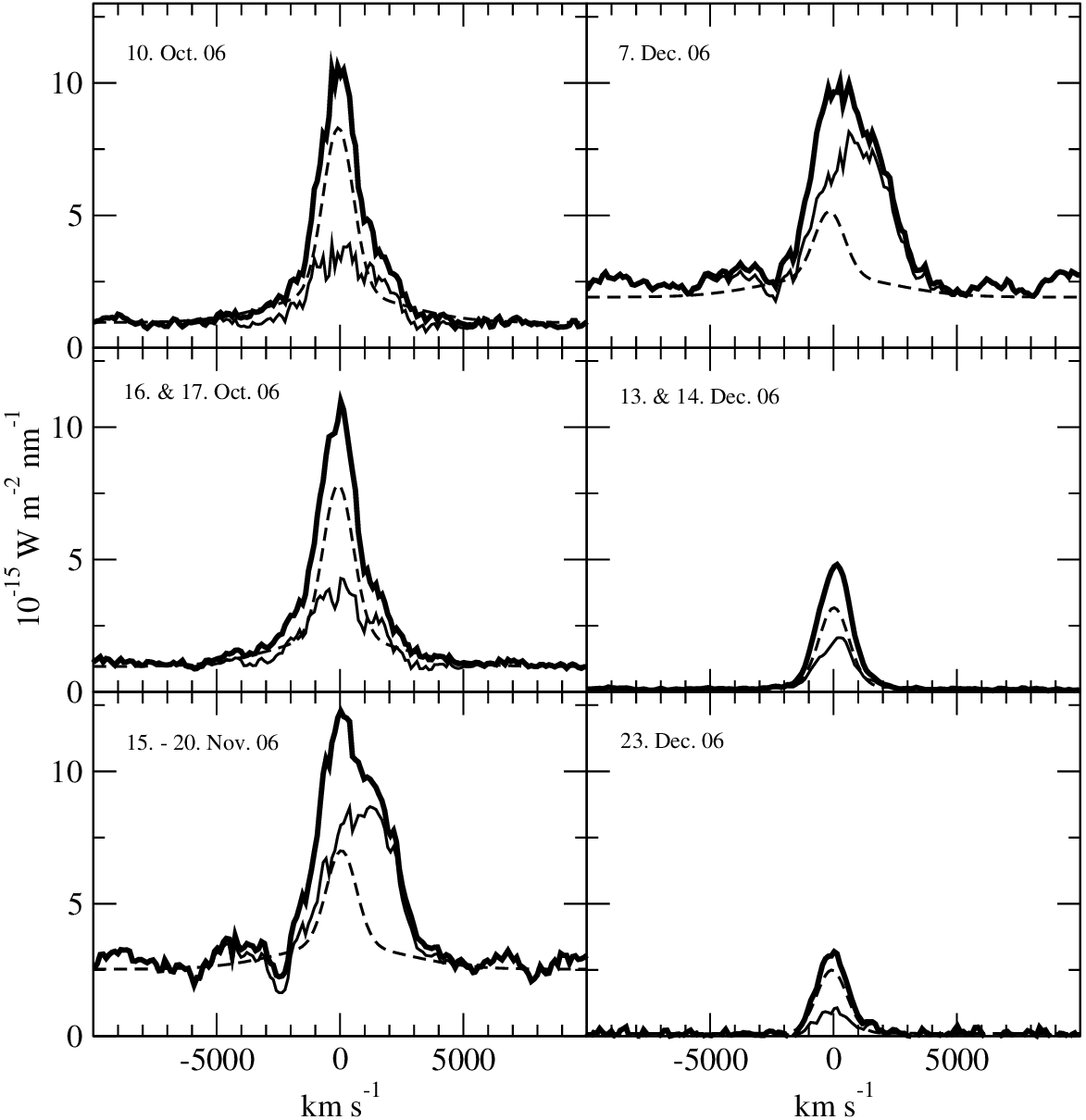}} \caption{The
H$_\alpha$ line illustrates the evolution of the different
structures. The total emission (thick line), the fitted emission of
the main outburst (dashed line) and the residual = component
appearing only during the bump (thin line). }\label{h-alpha}
\end{figure}

Due to the lower S/N in the other lines, we focus exclusively on the
remarkable line evolution using the strongest line, H$_\alpha$ (Fig.
\ref{h-alpha}). To gain insight into the nature of the bump event we
fit the narrow emission using a simple optically-thin shell model
arising from the main eruption. The bump event itself, far from
being near the optically-thin regime, cannot be modeled without
comprehensive radiative transfer calculations. Assuming an
instrument resolution folding of 0.6~nm, the narrow emission profile
perfectly fits that of two independent optically thin expanding
shells modeled after \citet{lamers} at the velocities reported by
\citet{siviero} and \citet{munari_b}. The thickness of the shells
are assumed to be negligible compared to their radii at this time.
An integration along line of sight results in a normalized line
profile $f(v;\lambda)$ of a single shell is as function of the
velocity $v$. The line profile for the detached shells expelled in
April 2006 is
\[g(v_1, v_2, a_0, r, c; \lambda) = a_0 \times \left({f(v_1;\lambda) + r\,\times\,f(v_1;\lambda)}\right) + c\]
where $v_1$ and $v_2$ are the expansion velocities of the shells,
$c$ the level of the continuum.  $a_0$ and $r$ are the fitting
parameters for the total intensity and for the intensity ratio. This
function was fitted to the data points $y_i$ between $\lambda_{\rm
MIN} = \lambda_0 - 10.5 \,{\rm nm}$ and $\lambda_{\rm MAX} =
\lambda_0 + 10.5\,$nm.
The velocities were fixed to those derived from the spectra of main
eruption by \citet{munari_b}; (1805 and 825 km
s$^{-1}$~respectively) to reduce the number of free parameters. With
the boundary condition that there has to be no absorption redshifted
from the maximum,
\[ [y_i - g(\lambda_i)] \ge c \qquad \qquad \forall \lambda \ge \lambda_0 = 656.3 {\rm nm}\]
and the continuum level $c$ derived outside the wavelength range we
used for the fit minimum of the logarithmic likelihood
\citep{statistik},
\[I(a_0, r) = - \sum_{\lambda_{\rm MIN}}^{\lambda_{\rm MAX}}{y_i \ln\left({g(a_0, r, c; \lambda_i) \over \sum{g(a_0, r, c; \lambda_i)}}\right)}\]
for the spectrum of October 10, 2006. The intensity ratio, $r$, was
0.78 on this date. As the main outburst ejecta were optically thin
by this time, the shape and intensity ratio can be assumed to be
constant for the other epochs.  This assumption also avoids
numerical instabilities from large numbers of free parameters. Note
that the phenomenological model cannot provide absolute values for
radii, densities, and physical conditions in these shells. The
absolute minimum of the statistical likelihood is outside the
physically-founded boundary; thus, a ``goodness of fit'' (e.g. by
$I(a_0, r) \le I_{MIN} + 0.5 \times n^2$, where $n$ gives the number
of $\sigma_{rms}$ requested) cannot be derived.

\begin{figure}[ht]
\centerline{\includegraphics[width=80mm]{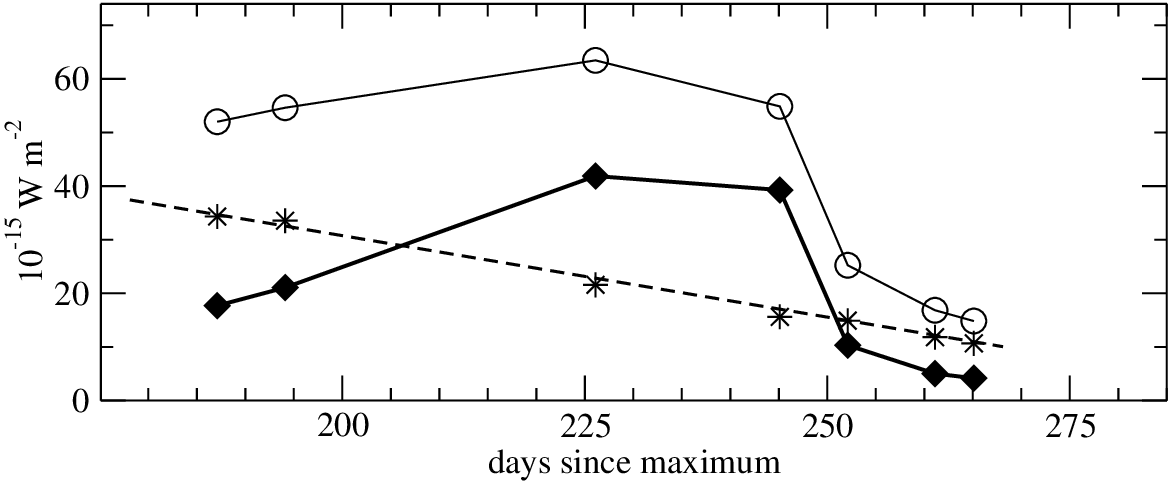}} \caption{The
flux evolution of the total H$_\alpha$ line flux (open circles), the
fitted emission of the main outburst (asterisk) and the residual
emission of the fast ejecta during the second maximum (filed
diamonds).}\label{h-alpha2}
\end{figure}

The decline of the fitted component was remarkably smooth and
homogenous, 1-2 \% per day (Fig. \ref{h-alpha2}), as is expected for
completely detached shells. Fitting the remaining structure after
subtracting the narrow emission model provides details on the bump
emission. The newly formed shell had a velocity of about
2\,600~km~s$^{-1}$ and appeared early in October 2006 ($\approx
180^{\rm d}$). Its emissivity peaked at the maximum of the
brightening.

\section{Discussion}
The derived ratio of $t_2/t_3$ and the decline rate of the first
60\,\,days and 4\fm0 in brightness fit perfectly the normal decline
described in \citet{kato_06,kato_07}.  In addition, this ratio also
resembles the prediction from a linear fit through the novae with
$t_2 < 20\fd0$ in Tab. 5.2 of \citet{warner}. The correlation of the
nova speed with the expansion velocities \citep{mclaughlin,warner}
also resembles that of a normal system. Thus, it seems reasonable to
use the samples in \citet{kato_06,kato_07} to estimate the white
dwarf mass $M_{\rm WD} = 1.2\pm0.1 M_\odot$ with $t_3$. Using the
curves of \citet{kato_1} leads to an even higher estimated mass.
Although the samples mentioned above have a very large scatter of
individual objects, they also give estimates for such mass for the
accretor. On the other hand, even if the early decline looks very
much like that of a normal object, one  has to be dubious whether
one is allowed to use such models for a nova with such a massive
secondary maximum. Thus, the results here have to be estimates. For
the proposed orbital period of 0\fd2070 \citep{goranskij_1} and the
mass of the accreting white dwarf of $M_{\rm WD} \approx
1.2$~M$_\odot$ the secondary has to be a late type $M_{\rm SE}\le
0.75~$M$_\odot$ K to M star.

We used \citet{Cox} to derive a temperature $T$ from stellar colors
and the evolution of the luminosity $L$ from the lightcurve and the
bolometric corrections. During the bump the temperature increased
slightly from 4\,250 to 4\,900~K and decreased again at maximum to
4\,250~K (see ($V-I_{\rm C}$) in Fig. \ref{lightcurve}). The
temperature derived from the early spectra \citep{early_spec} was
used as the temperature at maximum. Thereafter the evolution of the
radius of the ``quasi'' photosphere derived by $R \propto
T^{-2}\,L^{0.5}$ was obtained. The shrinking of the photosphere
stopped at day 60 and stayed constant at 15~\% of the radius of
maximum until day 170. In case of a homologous expansion this is
only possible with a significant increase of density. The fast
ejecta reached the photosphere around day 170. With a speed of
2\,200\,km\,s$^{-1}$ -- \ derived from our spectra and consistent
with the value measured in November 2006 by \citet{munari_b} -- and
the estimated radius above, this  bump material  was accelerated
about 45 to 60 days after maximum, assuming a negligibly small
radius of the accelerating region. Subtracting the extrapolated
power law decline and assuming that the bolometric correction
derived by the color is constant at $\approx$ 0\fm12 to 0\fm2, leads
to an estimate of the radiative energy in the event. About 40\% of
the total radiated energy of the object during the whole outburst is
contained in the bump. The high velocity of the fast ejecta during
this epoch also significantly increased the mechanical energy. Hence
we presume that the bump event contained a significant fraction of
the total outburst energy of the system.

The onset of massive dust formation, marked by the rapid decline,
the fast rise of the color (see Fig. \ref{lightcurve}), and the IR
excess \citep{ir_late} around day 250 occurred most likely in the
inner region only. Thus, the spectral components of the outer
optically-thin regions continued with a smooth decline, while the
fast component was obscured within a few days (see Fig.
\ref{h-alpha2}). The forbidden [OI] lines, which showed only the
narrow component throughout the whole event, were generated in the
outer region only. Although the obscured inner region, even seen
later in the IR excess \citep{ir_may07}, cannot directly heat the
outer shell, an effective energy transport changed the excitation
state quickly -- [OI] $\rightarrow$ [OII] $\rightarrow$ [OIII]
within 2 weeks.

The distance of $D \approx 7.5^{+3.0}_{-2.5}$~kpc places V2362~Cyg
at a galactocentric radius of about 11~kpc and a distance to the
plane of about 310~pc. As the direction marks the maximum of the
Galactic warp above the plane \citep{warp}, the real distance to the
disk is about twice this value. This coincides with the spiral arm
+II beyond the Perseus arm \citep{perseus}.

Further investigations of the ejecta of this unusual object and
collection of all the data of various observers from the early
decline until now is required to build the basis for a more
sophisticated theoretical modeling. We would like to encourage
modeling groups and would like to enliven a scientific discussion
about the nature of the second outburst of V2362~Cyg and its older
twin V1493~Aql.

\begin{acknowledgements} We thank the referee Greg Schwarz for the
extensive discussion improving the original manuscript. We
acknowledge the variable star observations from the AAVSO
International Database contributed by
observers worldwide and used in this research.\\

\end{acknowledgements}

%%\listofobjects
\end{document}